\documentclass{article}
\usepackage{CJKutf8}

\usepackage{amssymb,amsfonts,amsmath}
\usepackage{cite,enumerate,float,indentfirst}
\usepackage{color}

\def\be{\begin{eqnarray}}
\def\ee{\end{eqnarray}}
\def\nn{\nonumber}

\definecolor{red}{rgb}{1,0,0}
\definecolor{orange}{rgb}{1,0.5,0}
\definecolor{violet}{rgb}{0.7,0,1}

\hoffset= - 1.0in         

\begin{document}
\begin{CJK}{UTF8}{gbsn}
\title{\vspace{0.1cm}{\Large {\bf Differential expansion for link polynomials}\vspace{.2cm}}
\author{{\bf 白承铭 C.Bai$^a$}, \ {\bf 蒋剑剑 J.Jiang$^b$}, \ {\bf 梁晋婷 J.Liang$^a$},
\ {\bf A. Mironov$^{c,d,e,f}$}, \\
{\bf A. Morozov$^{d,e,f}$}, \ {\bf An. Morozov$^{d,e,f}$}, \ {\bf A. Sleptsov$^{d,e,f,g}$}
}
\date{ }
}

\maketitle
\end{CJK}

\vspace{-5.5cm}

\begin{center}
\hfill FIAN/TD-22/17\\
\hfill IITP/TH-26/17\\
\hfill ITEP/TH-16/17\\
\end{center}

\vspace{4.2cm}

\begin{center}

$^a$ {\small {\it Chern Institute of Mathematics and LPMC, Nankai University, Tianjin 300071, China}}\\
$^b$ {\small {\it School of Math \& Physics, Ningde Normal University, Ningde 352100, China}}\\
$^c$ {\small {\it Lebedev Physics Institute, Moscow 119991, Russia}}\\
$^d$ {\small {\it ITEP, Moscow 117218, Russia}}\\
$^e$ {\small {\it National Research Nuclear University MEPhI, Moscow 115409, Russia }}\\
$^f$ {\small {\it Institute for Information Transmission Problems, Moscow 127994, Russia}}\\
$^g$ {\small {\it Laboratory of Quantum Topology, Chelyabinsk State University, Chelyabinsk 454001, Russia }}
\end{center}

\vspace{1cm}

\begin{abstract}
The differential expansion is one of the key structures reflecting
group theory properties of colored knot polynomials,
which also becomes an important tool for evaluation of non-trivial
Racah matrices.
This makes highly desirable its extension from knots to links,
which, however, requires knowledge of the $6j$-symbols, at least, for
the simplest triples of non-coincident representations.
Based on the recent achievements in this direction, we conjecture
a shape of the differential expansion for symmetrically-colored links
and provide a set of examples.
Within this study, we use a special framing that is an unusual extension of the topological
framing from knots to links.
In the particular cases of Whitehead and Borromean rings links,
the differential expansions are different from the previously discovered.
\end{abstract}

\vspace{.5cm}

\section{Introduction}

Knot theory is currently one of the main applications of quantum field theory,
where non-perturbative results can be reliably derived and tested.
One of the subjects to study is the representation dependence of Wilson loop averages
(which, in four dimensions, distinguishes between the area and perimeter laws).
The $3d$ Chern-Simons theory \cite{CS} underlying knot theory is topological,
therefore, its observables can not depend on the metric data like lengths and areas,
still their dependence on representations is quite non-trivial.
Differential expansion \cite{Gukov}-\cite{DEdoublebraids}
is the simplest manifestation of such properties.
The goal of this paper is to extend the knowledge about differential expansion
from knots to links.
This is an essentially new story, because links consist of different components,
each in its own representation.
Calculation requires ${\cal R}$- and Racah matrices in channels with different
representations, which are not yet well studied and where some useful
methods like the eigenvalue conjecture \cite{eigenvalue} are not directly applicable.
Important for differential expansion is the choice of framing. For knots,
the best choice is the topological framing, when the ${\cal R}$-matrix is normalized in such a way that
it provides invariance with respect to the first Reidemeister move.
When the ${\cal R}$-matrix acts on a pair of different representations $R_1\otimes R_2$,
the first Reidemeister move is not applicable, and the topological framing is not
defined from the first principles. Even in the case of links, there is a distinguished {\it canonical framing} (or standard framing) \cite{Marino} suggested by M. Atiyah in \cite{At}. It turns out, however, that the differential expansion requires a bit different framing.

We briefly remind what the differential expansion is in sec.\ref{destr}
and formulate it for links of different kinds: made from unknots and
from non-trivial knots.
These conjectures are extracted from calculations of numerous examples,
which became possible due to advances in Racah calculus in \cite{Racahlinks,Racahlinks1}.

\section{Implications of representation theory}

The HOMFLY-PT polynomials \cite{HOMFLY-PT}
\be
\left.H_R^{\cal K}(A,q)\right|_{q=e^{\frac{2\pi i}{k+N}},\ A=q^N} =
\left<{\rm Tr}_R \ {\rm Pexp}\left(\oint_{\cal K} {\cal A}\right)\right>^{CS^{SU(N)}_k}
\ee
are characters of the loop algebra in Chern-Simons theory \cite{CS}
and inherit a lot of properties of characters of the ordinary groups.
In particular, the antisymmetric representation $[1^r]$ and
its  conjugate $[1^{N-r}]$ are isomorphic in $SU(N)$, thus,
the corresponding normalized HOMFLY-PT polynomials are the same in the topological framing:
\be
\frac{H_{[1^r]}}{D_{[1^r]}} - \frac{H_{[1^{N-r}]}}{D_{[1^{N-r}]}}\ \  \ \vdots \ \ \ \{A/q^N\}
\ee
where $\{x\} \equiv x-\frac{1}{x}$ and
$D_R(A,q) = \chi_R\left(p_k=\frac{\{Aq^k\}}{\{q^k\}}\right)$
are the values of Schur functions (characters) at the topological locus \cite{DMMSS,MMM2},
for $A=q^N$ they turn into dimensions of representations of $SU(N)$,
which are themselves invariant under conjugation.
Strictly speaking, the difference at the l.h.s. needs just to vanish at $A=q^N$,
the stronger statement with the factor $\{A/q^N\} \sim (A-q^N)(A^{-1}-q^{-N})$
is true only in the special (topological) framing.
Since the transposition of representations is equivalent to the change $q\longrightarrow -q^{-1}$
in knot polynomials, as a corollary, we get for the symmetric representations $R=[r]$
and $[N-r]$ (which are no longer conjugate) and for arbitrary $N$:
\be
\frac{H_{r}}{D_r}-\frac{H_{N-r}}{D_{N-r}}
\ \ \ \vdots \ \ \ \{Aq^N\}\{A/q\}
\label{Nrefl1}
\ee
The second factor at the r.h.s. reflects the triviality of reduced knot polynomials
in the Abelian case $N=1$, i.e. at $A=q$.
This again depends on the choice of the topological framing, otherwise the items at the l.h.s.
would be multiplied by different powers of $A$ and $q$ proportional to the writhe number,
and (\ref{Nrefl1}) would be violated.
The whole $\{A/q\}$ instead of just $(A-q)$ appears because the powers in $A$ and $q$
in all the terms of the knot polynomials have the same parity.

Our {\bf first claim} in this paper is that there exists a framing in which a similar
relation is true for links:
\be
\frac{ H_{r_1\,\ldots\,r_l}}{D_{r_1}\ldots D_{r_l}} \ -\
\frac{  H_{N-r_1\,\ldots\,N-r_l}}{D_{N-r_1}\ldots D_{N-r_l}}
\ \ \ \vdots \ \ \ \{Aq^N\}\{A/q\}
\label{Nrefll2}
\ee
where $l$ is the number of link components.

Our {\bf second claim} is that these simple relations have far-going consequences,
leading to a very powerful and restrictive representation of knot/link polynomials:
the differential expansion.
It works in an especially impressive way for knots in symmetric (and antisymmetric)
representations, and we begin with reminding this part of the story.
For an even more impressive (though far more involved)
lift to rectangular representations, see \cite{Rrect}.
The main part of the present letter is another extension:
still symmetric representations, but for links.
In principle, in this case, it is very useful to distinguish the link made
from unknot components from those, where components are knotted themselves,
though this issue is not that important as compared with the very fact of existence of
a special framing which provides a {\bf differential expansion for links}.

\section{Differential expansion for knots in symmetric representations \label{destr}}

As a corollary of (\ref{Nrefl1}), we obtain the following formula
\cite{IMMMfe}-\cite{DEdoublebraids},
which we call {\it differential expansion}:
\be
\boxed{
{\cal H}_r^{\cal K} =
\frac{H_r^{\cal K}}{D_r} = 1 + \sum_{s=1}^r \frac{[r]!}{[s]![r-s]!}\cdot F_s^{\cal K}(A,q)\cdot
\{A/q\}\cdot \prod_{j=0}^{s-1}\{Aq^{r+j}\}
}
\label{deknots}
\ee
Indeed, ${\cal H}_0 = 1$, and (\ref{Nrefl1}) at $N=r$ implies that
\be
{\cal H}_r = 1 + \{Aq^r\}\{A/q\}\cdot F_r
\ee
with some $F_r(q,A)$.
At $N=r+1$, one gets additionally
\be
\frac{{\cal H}_r-{\cal H}_1}{\{A/q\}} = \{Aq^r\}F_r - \{Aq\} F_1 \ \ \vdots  \ \ \{Aq^{r+1}\} \ \
\Longrightarrow \ \  F_r = [r] \cdot F_1 + \{Aq^{r+1}\}\tilde F_r
\ee
After this substitution at $N=r+2$,
\be
\frac{{\cal H}_r-{\cal H}_2}{\{A/q\}}
= \Big(\underbrace{[r]\{Aq^r\}-[2]\{Aq^2\}}_{[r-2]\{Aq^{r+2}\}}\Big) F_1
+ \{Aq^r\}\{Aq^{r+1}\}\tilde F_r - \{Aq^3\}\{Aq^2\} \tilde F_2 \ \vdots  \ \{Aq^{r+2}\} \ \
\Longrightarrow \nn \\ \Longrightarrow \ \   \tilde F_r =
\frac{[r][r-1]}{[2]} \tilde F_2 + \{Aq^{r+2}\}\tilde{\tilde F}_r
\ee
where the coefficient $\frac{[r][r-1]}{[2]}$ is the value of $\{Aq^3\}\{Aq^2\}$ at
$A=q^{-r-2}$.
Repeating the same procedure for higher and higher $N$ and adjusting the notation, one
gets (\ref{deknots}).

\section{Defects and additional factorization}

For the fixed $N$ and $q=e^\hbar$, $\{Aq^n\} = \{q^{N+n}\}\sim \hbar$.
The functions in this case are of the order $F_s \sim \hbar^{s-1}$.
Moreover, for each knot, there is a special integer-valued characteristic
$\delta^{\cal K}$ \cite{konodef} such that, for large enough $s$, there is an additional factorization:
\be
F_s^{\cal K}(A,q) = G_s^{\cal K}(A,q)\cdot \prod_{j=0}^{\nu^{\cal K}-1}\{Aq^j\}
\ee
so that $G_s\sim \hbar^{s-\nu^{\cal K}}$ with
\be
\nu^{\cal K} = {\rm entier}\left(\frac{s-1}{\delta^{\cal K}+1}\right)
\ee
The parameter $\delta^{\cal K}$ is called {\it defect} of the differential expansion,
it is equal to zero for twist knots, and to $k-1$ for the 2-strand torus knots
$[2,2k+1]$.
The functions $F_s^{\cal K}$ for the particular knot are not all independent,
but exact relations between them are not known yet (see, however, a recent \cite{Sulkquiv} for a suggestion to parameterize
the knot dependence of $\{F_s\}$ by auxiliary quivers).
At the same time, they are not sufficient to characterize the knot completely:
the mutant pairs are separated only by the HOMFLY-PT polynomials in non-symmetric
representations.
The differential expansions for them also exist, but are far more complicated,
see \cite{DEnonsym,Rama1,DEdoublebraids} for the first steps in this direction.

\section{Differential expansion for two-component links}

Now we can proceed to links.
Again, as a particular case of (\ref{Nrefll2}), for 2-component links
\be
{\cal H}_{r_1,r_2} - {\cal H}_{N-r_1,N-r_2}\ \ \ \vdots \ \ \ \{Aq^N\}\{A/q\}
\label{Nrefl2}
\ee
and just by the same arguments as in sec.\ref{destr} we get for
{\bf links made from two unknots}:
\be
\boxed{
{\cal H}_{r_1, r_2}^{\cal L} =
\frac{H_{r_1, r_2}^{\cal L}}{D_{r_1}\cdot D_{r_2}} = 1 +
\sum_{s=1}^{r_2} \frac{[r_2]!}{[s]![r_2-s]!} \cdot F_{s\,|\,r_1-r_2}^{\cal L}(A,q)\cdot
\{A/q\}\cdot \prod_{j=0}^{s-1}\frac{\{Aq^{r_1+j}\}}{\{Aq^j\}}
}
\label{unknotlinks}
\ee
only now we are left with an additional uncontrolled dependence of
$r_1-r_2$ in $F_{s\,|\,r_1-r_2}\sim \hbar^{s-1}$. Here $r_1\ge r_2$.

Indeed, putting $N=r_1$
and using the fact that ${\cal H}^{\cal L}_{r,0} = 1$,
we get the simplest precursor of (\ref{unknotlinks}) in the form
\be
{\cal H}^{\cal L}_{r_1,r_2 } -1 =
{\cal F}^{\cal L}_{ r_1,r_2}\cdot \{Aq^{r_1}\}\{A/q\}
\label{2link1}
\ee
Next, at $N=r_1+1$, from
\be
\frac{{\cal H}^{\cal L}_{r_1,r_2} - {\cal H}^{\cal L}_{r_1-r_2+1,1}}{\{A/q\}}
={\cal F}_{r_1,r_2}^{\cal L}\cdot\{Aq^{r_1}\}
-{\cal F}^{\cal L}_{r_1-r_2+1,1}\cdot\{Aq^{r_1-r_2+1}\}
\ \ \vdots \ \ \{Aq^{r_1+1}\}
\ee
we get
\be
{\cal F}^{\cal L}_{r_1,r_2} = [r_2]\cdot {\cal F}^{\cal L}_{r_1-r_2+1,1} + \tilde{\cal F}^{\cal L}_{r_1,r_2}\cdot \{Aq^{r_1+1}\}
\ee
because $[r_2]\{Aq^{r_1}\} - \{Aq^{r_1-r_2+1}\}=[r_2-1]\{Aq^{r_1+1}\}$.
In fact, ${\cal F}_{r_1,r_2}$ is a polynomial divided by $\prod_{j=0}^{r_2-1}\{Aq^j\}$,
in particular, ${\cal F}_{r_1-r_2+1,1}=\frac{1}{\{A\}}\cdot F_{1|r_1-r_2}(A,q)$,
where the new function is already a polynomial, and its notation is made better adjusted
to the needs of the differential expansion.
Taking this into account, we obtain the first term of the expansion (\ref{unknotlinks}):
\be
{\cal H}^{\cal L}_{r_1,r_2} = 1
+ [r_2]\cdot F^{\cal L}_{1|r_1-r_2}\cdot \frac{\{Aq^{r_1}\}\{A/q\}}{\{A\}}
+ O\Big(\{Aq^{r_1+1}\}\{Aq^{r_1}\}\Big)
\label{2link2}
\ee
Proceeding further to higher $N-r_1$, we obtain the entire (\ref{unknotlinks}).

In the case arbitrary two-component links
(where particular components can be themselves knotted),
it is necessary to substitute unity in the first term at the r.h.s. of (\ref{unknotlinks})
by the product of the corresponding reduced knot polynomials:
\be
\boxed{
{\cal H}_{r_1,  r_2}^{\cal L} \ - \
{\cal H}_{r_1, \emptyset}^{\cal L}\cdot{\cal H}_{\emptyset,r_2}^{\cal L}
= \sum_{s=1}^{r_2 } \frac{[r_2]!}{[s]![r_2-s]!}  \cdot
F_{s\,|\,r_1-r_2 }^{\cal L}(A,q)\cdot
\{A/q\}\cdot \prod_{j=0}^{s-1}\frac{\{Aq^{r_1+j}\}}{\{Aq^j\}}
}
\label{DElink}
\ee
and the r.h.s. can be also restructured to separate the contributions
of constituent knots from the link itself.

Formula (\ref{unknotlinks}) can be easily checked in the case of torus links,
where the HOMFLY-PT invariant in arbitrary representation is immediately
provided by the Rosso-Jones formula \cite{RJ,LZ}.
This does not lead to a proof for generic links, but fixes all the
ambiguities, which can afterwards be tested in any other example.
Unfortunately, within the torus family all the links are made from identical torus knots
(not obligatory {\it un}knots), which makes the knot invariant symmetric w.r.t. permutations of the representations of different components, and this case is not enough to provide convincing evidence.

To better study linking of non-trivial knots,
one needs some more powerful technique than the Rosso-Jones formula.
In this paper, we apply the mixing-matrix and arborescent calculus of \cite{MMM2,modRT1,modRT2}
and \cite{arbor,Rama1}, the far-going generalizations of the Reshetikhin-Turaev approach
\cite{RT},
which, in turn, requires knowledge of the Racah matrices
\be
{\cal U}:\ \ \Big((R_1\otimes R_2)\otimes R_3\longrightarrow Q\Big) \ \longrightarrow\
\Big(R_1\otimes (R_2 \otimes R_3)\longrightarrow Q\Big)
\ee
for non-coinciding representations $R_1,R_2,R_3$, which were calculated
in \cite{Racahlinks1}.

\section{Extension to three-component links}

For links with {\bf more components}, there are additional dependencies on $r_1-r_k$.
As in the case of knots, there can also occur an additional factorization for
$s\geq\delta^{\cal L}+2$.

Let us consider a three-component link ${\cal L}_3$ made from three unknots.
In this case, one can use (\ref{Nrefll2}) in the form
\be
{\cal H}^{{\cal L}_3}_{r_1,r_2,r_3} - {\cal H}^{{\cal L}_3}_{N-r_1,N-r_2,N-r_3} \ \vdots \
\{Aq^N\}\{A/q\}
\ee
with the order of representations $r_1\geq r_2\geq r_3$.
Putting $N=r_1$, we get the analogue of (\ref{2link1}):
\be
{\cal H}^{{\cal L}_3}_{r_1,r_2,r_3} - {\cal H}^{{\cal L}_2}_{r_1-r_3,r_1-r_2} =
{\cal F}^{{\cal L}_3}_{r_1,r_2,r_3}\cdot \{Aq^{r_1}\}\{A/q\}
\ee
where ${\cal L}_2$ denotes the link with one removed unknot.
Now the second item at the l.h.s. is not just unity, but the two-component link.

Next, at $N=r_1+1$, from
\be
\frac{{\cal H}^{{\cal L}_3}_{r_1,r_2,r_3} - {\cal H}^{{\cal L}_3}_{r_1-r_3+1,r_1-r_2+1,1}}{\{A/q\}}
= {\cal F}^{{\cal L}_3}_{r_1,r_2,r_3}\cdot \{Aq^{r_1}\} -
{\cal F}^{{\cal L}_3}_{r_1-r_3+1,r_1-r_2+1,1}\cdot \{Aq^{r_1-r_3+1}\}
\ \ \ \vdots \ \ \  \{Aq^{r_1+1}\}
\ee
we obtain:
\be
{\cal F}^{{\cal L}_3}_{r_1,r_2,r_3} = [r_3]\cdot {\cal F}^{{\cal L}_3}_{r_1-r_3+1,r_1-r_2+1,1}
+ \tilde{\cal F}^{{\cal L}_3}_{r_1,r_2,r_3}\cdot \{Aq^{r_1+1}\}
\ee
In fact, ${\cal F}_{r_1,r_2,r_3}$ is a polynomial divided by $\prod_{j=0}^{r_2-1}\{Aq^j\}$,
in particular, ${\cal F}_{r_1-r_2+1,1}=\frac{1}{\{A\}}\cdot {\cal F}_{1|r_1-r_2}(A,q)$,
where the new function is already a polynomial, and its notation is made better adjusted
to the needs of the differential expansion.
Taking this into account, we obtain the first term of the expansion:
\be
{\cal H}^{{\cal L}_3}_{r_1,r_2,r_3} - {\cal H}^{{\cal L}_2}_{r_1-r_3,r_1-r_2}
= [r_3]\cdot {\cal F}^{{\cal L}_3}_{1|r_1-r_2,r_1-r_3}\cdot
\frac{\{Aq^{r_1}\}\{A/q\}}{\{A\}} + O\Big(\{Aq^{r_1+1}\}\{Aq^{r_1}\}\Big)
\ee

\section{On the choice of framing}

Eq.(\ref{unknotlinks}) can seem to be in apparent contradiction already
with the well-known result for Hopf link (torus $[2,2]$) \cite{Vafa,arthboro}:
\be
\frac{h^{\rm Hopf}_{r_1, r_2}(A,q)}{D_{r_1}\cdot D_{r_2}}
= 1 + \sum_{k=1}^{{\rm min}(r_1,r_2)}
(-)^kA^{-k}q^{-k(r_1+r_2)+\frac{k(k+3)}{2}}
\prod_{j=0}^{k-1} \frac{\{q^{r_1-j}\}\{q^{r_2-j}\}}{\{Aq^j\}}
\label{arthHopf}
\ee
The reason for this discrepancy is in different framings of HOMFLY-PT
invariants in these cases.

Indeed, as we already mentioned, there is the canonical framing \cite{At,Marino} defined by the requirement of zero self-linking number, which, in particular, implies that the linear in $\hbar$ term in the expansion of the link invariant vanishes. In the case of knots, this framing coincides with the topological one. In the case of links, in this framing, there is also an additional topological factor $q^{2\sum_{i>j}|R_i||R_j| Lk_{ij}/N}$, where $Lk_{ij}$ is the linking number of the $i$-th and $j$-th components of the link and $|R_i|$ is the number of boxes of the Young diagram corresponding to the representation $R_i$. This factor is in charge of the difference between $U(N)$ and $SU(N)$ invariants and is often omitted (see, e.g., \cite{LM,LZ}), as we do (this makes the link invariant a rational function of $A$ and $q$). In particular, formula (\ref{arthHopf}) as well as the link invariants for the Whitehead (\ref{WH}) and for the Borromean rings (\ref{B}) are written in the canonical framing.

Let us first consider the torus link case. The basic for calculations in torus family is the Rosso-Jones formula,
\be
h^{[m,n]}_{_{R_1, \ldots, R_m}}(A,q) \ \ = q^{\frac{2n}{m}(\varkappa_{R_1}+\ldots+\varkappa_{R_m})}
\sum_{Q\vdash |R_1|+\ldots +|R_m|} C_Q\cdot D_Q(A,q)\cdot q^{\frac{2n}{m}\varkappa_Q}
\label{RJf}
\ee
where the sum goes over all the Young diagrams $Q$ of the size
$|R_1|+\ldots|R_m|$ and $\varkappa_Q =\sum_{(i,j)\in Q}(i-j)$.
The link $[m,n]$ is a knot for coprime $n$ and $m$, then all $R_i$ should be the same.
If the biggest common divisor of $m$ and $n$ is $l$, then we get an $l$-component link
made from the knots $[m/l,n/l]$, and there can be $k$ different representations:
$R_{pl+r} = R^{(r)}$ with $r=1,\ldots,l$.
The coefficients $C_Q$ are provided by the Adams decomposition of Schur functions:
\be
\prod_{i=1}^l \chi_{_{R^{(i)}}}\{p_{mk/l}\} \ \
=\sum_{Q \vdash \frac{m}{l}\left(|R^{(1)}|+\ldots +|R^{(l)}\right)} C_Q\chi_Q\{p_k\}
\ee
For 2-strand links ($m=2$, $n=2k$, $l=2$) and for symmetric representations $R^{(i)}=[r_i]$
(\ref{RJf}) turns into a much simpler formula, involving only two-line Young diagrams:
\be
h_{r_1, r_2}^{[2,2k]}(A,q)= q^{2k(\varkappa_{r_1}+\varkappa_{r_2})}
\sum_{i=0}^{r_2} D_{[r_1+r_2-i,i]}(A,q)\cdot q^{-2k\varkappa_{[r_1+r_2-i,i]}}
\ee
The Rosso-Jones formula is given in the canonical framing. This means that all ${\cal R}$-matrices are normalized in the topological framing for coinciding representations. In the case of link, one has, however, two types of ${\cal R}$-matrices: those acting on the pair of representations from the same component of the link, and, from different components of the link. The ${\cal R}$-matrices of the first type always act on the pair of coinciding representations, and one can naturally choose them to be in topological framing. At the same time, the ${\cal R}$-matrices of the second type could have different normalization.

In fact, one can calculate what normalization of the ${\cal R}$-matrix does not change the HOMFLY-PT invariant after conjugation, which is necessary for (\ref{Nrefll2}) to be correct. The answer is that one has to normalize the ${\cal R}$-matrices of the second type with an additional factor $\Big(Aq^{{\rm max}(r_i,r_j)-1}\Big)^{\,n\cdot {\rm min}(r_i,r_j)}$ (in fact, one can interchange $max$ and $min$ in this formula) for each crossing of $i$-th and $j$-th link components so that the link invariants satisfying (\ref{Nrefll2}) are related to those in the canonical framing via
\be\label{topfr}
\boxed{
{\cal H}_{r_1,\ldots,r_l}=\prod_{i>j}\Big(Aq^{{\rm max}(r_i,r_j)-1}\Big)^{2Lk_{ij}\cdot {\rm min}(r_i,r_j)}{h_{r_1,\ldots,r_l}\over D_{r_1}\ldots D_{r_l}}
}
\ee
For $r_1=r_2$, this factor reproduces the usual ratio
$A^{|R|}q^{2\varkappa_R}$ between the vertical and topological
framings. In other words, one has to take ${\cal R}$-matrices of the first type in the topological framing, and the ${\cal R}$-matrices of the second type in the vertical one. We call this framing {\it differential}.

Coming back to the difference between (\ref{unknotlinks}) and (\ref{arthHopf}), the former is true at the differential framing, while the latter one, at
the canonical framing. More concretely, for the Hopf link, (\ref{unknotlinks}) takes the form
\be
{\cal H}^{[2,2]}_{r_1, r_2} = 1 + \sum_{s=1}^{r_2} \frac{[r_2]!}{[s]![r_2-s]!}\cdot
A^s\cdot q^{\frac{s(s-1)}{2}}\cdot q^{s(r_1-r_2)}\cdot \frac{\{A/q\}}{\{Aq^{s-1}\}} \prod_{j=0}^{s-1}  \{Aq^{r_1+j}\}
\label{DEhopf}
\ee
The two other results from \cite{arthboro} in the canonical framing, for the Whitehead
\be\label{WH}
{\cal H}^{\rm WH}_{r_1, r_2}(A,q)
= 1 + \sum_{k=1}^{{\rm min}(r_1,r_2)}
 A^{-k}q^{-\frac{k(k-1)}{2}}\frac{\{A/q\}}{\{Aq^{k-1}\}}
\prod_{j=0}^{k-1} \frac{\{Aq^{r_1+j}\}\{Aq^{r_2+j}\}}{\{Aq^{k+j}\}}\{q^{r_1-j}\}\{q^{r_2-j}\}
\ee
and for the Borromean rings
\be\label{B}
{\cal H}^{\rm B}_{r_1, r_2, r_3}(A,q)
= 1 + \!\!\!\!\!\!\!\! \sum_{k=1}^{{\rm min}(r_1,r_2,r_3)}\!\!\!\!\!\!\!
 (-)^k\{q\}^k[k]! \frac{\{Aq^{k-2}\}! }{\{Aq^{2k-1}\}! }
\prod_{j=0}^{k-1} \{Aq^{r_1+j}\}\{Aq^{r_2+j}\}\{Aq^{r_3+j}\}\{q^{r_1-j}\}\{q^{r_2-j}\}\{q^{r_3-j}\}
\ee
are already in the form (\ref{unknotlinks}), so, in these two cases,
there is no need to perform any transformation like (\ref{topfr}),
required in the case of torus links.
This is not a big surprise: the Borromean rings are actually framing-independent, and the differential framing coincides with the canonical one for
the Whitehead because of the zero linking number.

\begin{figure}
\begin{picture}(360,150)(-240,-50)
\qbezier(-160,0)(-160,10)(-150,17)
\qbezier(-160,0)(-160,-10)(-150,-17)
\qbezier(-150,17)(-140,25)(-132,19)
\qbezier(-150,-17)(-140,-25)(-130,-16)
\qbezier(-130,-16)(-110,0)(-128,16)
\qbezier(-101,0)(-101,10)(-111,17)
\qbezier(-101,0)(-101,-10)(-111,-17)
\qbezier(-111,-17)(-121,-25)(-129,-19)
\qbezier(-111,17)(-121,25)(-131,16)
\qbezier(-131,16)(-151,0)(-133,-16)
\put(-20,-20){\line(1,1){40}}
\put(2,-2){\line(1,-1){22}}
\put(-2,2){\line(-1,1){22}}
\qbezier(22,22)(44,0)(24,-20)
\qbezier(-22,-22)(-44,0)(-24,20)
\qbezier(22,22)(0,44)(-20,24)
\qbezier(-22,-22)(0,-44)(20,-24)
\qbezier(24,24)(45,45)(65,15)
\qbezier(24,-24)(45,-45)(65,-15)
\qbezier(65,15)(73,0)(65,-15)
\qbezier(-24,24)(-45,45)(-65,15)
\qbezier(-24,-24)(-45,-45)(-65,-15)
\qbezier(-65,15)(-73,0)(-65,-15)
\qbezier(143,3)(155,19)(142,31)
\qbezier(140,0)(130,-10)(115,2)
\qbezier(138,35)(124,46)(112,30)
\qbezier(115,2)(100,12)(112,30)
\qbezier(145,16)(127,18)(120,2)
\qbezier(151,15)(163,13)(166,-6)
\qbezier(166,-6)(168,-30)(142,-30)
\qbezier(118,-2)(113,-30)(142,-30)
\qbezier(134,13)(141,-6)(163,0)
\qbezier(133,17)(130,26)(144,36)
\qbezier(144,36)(158,46)(170,32)
\qbezier(167,2)(187,9)(170,32)
\end{picture}
\caption{{\footnotesize These are three simplest links.
The leftmost figure shows the two-component Hopf link.
The one in the middle is the two-component Whitehead link and the rightmost link is the three-component Borromean rings link}}
\end{figure}
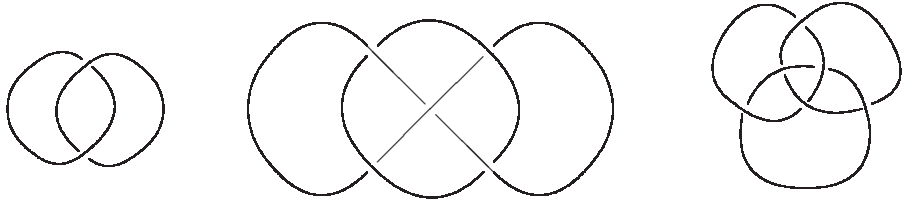

An advantage of (\ref{unknotlinks}) is that it respects
the group theory properties like (\ref{Nrefl1})
and (\ref{Nrefll2}) and thus is true in full generality in application to all links,
while expansions like (\ref{arthHopf}) exist in particular cases and sometime
look more like a miracle.

\section{Examples of differential expansion (\ref{unknotlinks}) }

In this section, we write down explicit examples of various links that demonstrate formulas (\ref{unknotlinks}), (\ref{Nrefl2}).

\subsection{2-strand torus links}

\paragraph{Hopf link $[2,2]$ }
The concrete values of the functions $F_{s|r_1-r_2}$ for the Hopf link are
\be
F^{[2,2]}_{s,r_1-r_2} = G^{[2,2]}_{s,r_1-r_2} \cdot{\prod_{j=0}^{s-2}\{Aq^j\}} =
A^s \cdot q^{\frac{s(s-1)}{2}} \cdot q^{s(r_1-r_2)}\cdot {\prod_{j=0}^{s-2}\{Aq^j\}}
\ee
In this particular case, they are further factorized, because the {\it defect}
of the Hopf link is zero.

\paragraph{Link $[4,2]$}

\be
\begin{array}{l}
F^{[4,2]}_{1,r_1-r_2}
= A\cdot q^{r_1-r_2}\cdot\left(1+\left(A^2(q^{-2}-1+q^{2(r_1-r_2+1)}\right)\right)
\nn\\
F^{[4,2]}_{2,r_1-r_2} = A \cdot q^{2(r_1-r_2)}\cdot \left(   [2]A^6\{q\}^2   +
q\{A\}\Big(A^5\cdot q^{4(r_1-r_2+2)} - [2] A^3(A^2\{q\}-q^{-1})\cdot q^{2(r_1-r_2+2)}
-[2]A^3\{q\}+A\Big)\right) \nn \\
\ldots
\end{array}
\ee
In this case the defect is 1, thus, an additional factorization occurs beginning from
the third function $F_s$.

\paragraph{Generic 2-strand torus links $[2k,2]$}

\be
\begin{array}{l}
F^{[2k,2]}_{1,1}(A,q)
= \sum_{j=0}^{k-1}\left(\frac{[2]}{[3]} q (Aq^2)^{2j} + \frac{1}{[3]}q^{-2} (A/q)^{2j}\right)
= \frac{1}{[3]}\left(q[2]\frac{(Aq^2)^{2k}-1}{(Aq^2)^2-1}
+ q^{-2}\frac{(A/q)^{2k}-1}{(A/q)^2-1}\right)\nn\\
\ldots
\end{array}
\ee
Defect is $k-1$ and the factorization appears beginning from $F_{2k-1}$.

\subsection{Whitehead link L5a1 }

As already noted, eq.(\ref{WH}) is already in the form very similar to (\ref{unknotlinks}),
but not quite.
To see the difference, let us look at the two simplest cases of $r_2=1$ and $r_2=2$:
\be
{\cal H}^{\rm W}_{r_1, 1}(A,q) \ \stackrel{(\ref{WH})}{=}\  1 +
&\underbrace{\frac{\{q^{r_1}\}\{q\}}{A}}
\cdot \frac{\{Aq^{r_1}\}\{A/q\}}{\{A\}}  \nn \\
{\cal H}^{\rm W}_{r_1, 1}(A,q) \ \stackrel{(\ref{unknotlinks})}{=}\  1 +
& {F^{W}_{1|r_1-1}}\cdot
\frac{\{Aq^{r_1}\}\{A/q\}}{\{A\}}
\label{WH1}
\ee
and
\be
{\cal H}^{\rm W}_{r_1, 2}(A,q) \ \stackrel{(\ref{WH})}{=}\ 1 +
&\frac{\{q^{r_1}\}\{q^2\}}{A}\cdot\frac{\{Aq^2\}}{ \{Aq\}} \cdot
\frac{\{Aq^{r_1}\}\{A/q\}}{\{A\}}\ +
&\frac{\{q^{r_1}\}\{q^{r_1-1}\}\{q^2\}\{q\} }{q\,A^2 } \cdot
\frac{\{Aq^{r_1}\}\{Aq^{r_1+1}\}\{A/q\}}{\{A\}} \nn \\
{\cal H}^{\rm W}_{r_1, 2}(A,q) \ \stackrel{(\ref{unknotlinks})}{=}\ 1 +
&[2]\cdot {F^{W}_{1|r_1-2}}\cdot
\frac{\{Aq^{r_1}\}\{A/q\}}{\{A\}} \ +
&{F^{W}_{2|r_1-2}}\cdot
\frac{\{Aq^{r_1}\}\{Aq^{r_1+1}\}\{A/q\}}{\{A\}\{Aq\}} \nn \\
&[2]\cdot \frac{\{q^{r_1-1}\}\{q\}}{\{A\}}\cdot
\frac{\{Aq^{r_1}\}\{A/q\}}{\{A\}} \ \
\ee
where, in the last line, we substituted $F^{W}_{1|r_1-2}$ from (\ref{WH1}).
One can substitute $[2]\{q\}=\{q^2\}$, still in the linear term there is a mismatch
in   $\frac{\{q^{r_1}\}\{Aq^2\}}{\{Aq\}}-\{q^{r_1-1}\} = \frac{\{q\}\{Aq^{r_1+1}\}}{\{Aq\}}$,
which has the right structure to be captured into $F^{W}_{2|r_1-2}$,
but makes it non-factorized.
In result, we obtain somewhat sophisticated expressions:
\be
\begin{array}{l}
F^{W}_{1|r} = \frac{[r+1]\{q\}^2}{A} \\
F^{W}_{2|r} = \frac{[2]\{q\}^2}{A^3}\Big(1+ (A^2-1)(q^{2r+2}-q^{-2}+q^{-2r-4})\Big) \\
F^{W}_{3|r} = \frac{[2][3][r+2]\{q\}^4\{A\}}{A^4}\Big(  A^2 (q^{2r+2}+q^{-2}-q^{-4}+q^{-2r-6}) - q^{-4}\{q\}^2[r+3][r+1]\Big) \\
F^{W}_{4|r} = \frac{[2][3][4][r+2]\{q\}^4\{A\}}{A^6}\Big(
A^4(q^{4r+7}-q^{2r-1}+q^{-7}-q^{-2r-11}-q^{-4r-13}) - \nn \\
- A^2\{q\}^2[r+3][r+2](q^{2r}+q^{2r-2}+q^{-6}-q^{-8}-q^{-10}+q^{-2r-10}+q^{-2r-12})
+q^{-9}\{q\}^4[r+4][r+3][r+2][r+1]
\Big)
\\
\ldots
\end{array}
\ee
The additional factors of $\{A\}$ in the third and forth line appear because the defect
of W link is $1$, the fifth and sixth lines will contain factors $\{A\}\{Aq\}$
and so on.
Comparison of these relatively sophisticated formulas with the much simpler (\ref{WH})
can imply that the structure of differential expansion
(\ref{unknotlinks}) can still be better tuned to make the functions $F$ simpler.

\subsection{Single-ring trefoil necklace}

Another instructive example is provided by the {\it neclaces},
composite links arising when the rings (unknots) are hanged on a knot ${\cal K}$.
In this case, the reduced link polynomials are
just products of those for the constituents.
For instance, when there is just one such a ring attached to a knot ${\cal K}$
the HOMFLY-PT for the corresponding two-component link is just the product
involving the Hopf link (\ref{DEhopf})
\be
{\cal H}_{r_1,r_2}^{{\cal K}\wedge 1} = {\cal H}^{\cal K}_{r_1}\cdot {\cal H}^{[2,2]}_{r_1,r_2}
\ee
and (\ref{Nrefll2}) is trivially satisfied, because it is true for the components
of the product.

Consider the simplest case of the link when $K$=trefoil. Since it is not a prime link, it has no an individual name in \cite{katlas}.
As soon as one of the components is a non-trivial knot,
we should prefer (\ref{DElink}) over (\ref{unknotlinks}). Then, the coefficients can be immediately read off from (\ref{DEhopf})

\begin{picture}(300,80)(-350,0)
\qbezier(0,20)(0,50)(30,50)
\qbezier(0,20)(0,-10)(38,-28)
\qbezier(42,50)(75,50)(55,0)
\qbezier(46,-31)(60,-37)(78,30)
\qbezier(5,-10)(0,-50)(25,-50)
\qbezier(7.5,0)(25,60)(70,60)
\qbezier(25,-50)(35,-50)(55,0)
\qbezier(70,60)(85,60)(80,38)
\qbezier(85,33)(70,33)(70,55)
\qbezier(70,65)(70,75)(85,75)
\qbezier(85,75)(100,75)(100,60)
\qbezier(85,33)(100,33)(100,60)
\end{picture}

\vspace{-3cm}

\begin{flushleft}
\parbox{10cm}{
\be
{\cal H}_{r,1}^{3_1\wedge 1}-{\cal H}_r^{3_1}={\cal H}_r^{3_1}\Big({\cal H}_{r,1}^{[2,2]}-1\Big)=F_{1|r}{\{A/q\}\{Aq^r\}\over\{A\}}\Longrightarrow
F_{1|r-1}^{{\cal K}\wedge 1}=Aq^{r-1}{\cal H}_r^{3_1}
\nn
\ee
Similarly,
$$
F_{2|r-2}^{{\cal K}\wedge 1}=A^2q^{2r-3}\{A\}{\cal H}_r^{3_1}+[2]Aq^{r-2}{\{Aq\}\over\{Aq^{r+1}\}}\Big({\cal H}_r^{3_1}-{\cal H}_{r-1}^{3_1}\Big)
$$
$$
F_{3|r-3}^{{\cal K}\wedge 1}=A^3q^{3r-6}\{A\}\{Aq\}{\cal H}_r^{3_1}+[3]A^2q^{2r-5}{\{A\}\{Aq^2\}\over\{Aq^{r+2}\}}\Big({\cal H}_r^{3_1}-{\cal H}_{r-1}^{3_1}\Big)
+
$$
$$
+[3]Aq^{r-3}{\{Aq\}\{Aq^2\}\over\{Aq^{r+1}\}\{Aq^{r+2}\}}\Big({\cal H}_r^{3_1}-{\cal H}_{r-2}^{3_1}\Big)-[2][3]Aq^{r-3}{\{Aq\}\{Aq^2\}\over\{Aq^{r}\}\{Aq^{r+2}\}}
\Big({\cal H}_{r-1}^{3_1}-{\cal H}_{r-2}^{3_1}\Big)
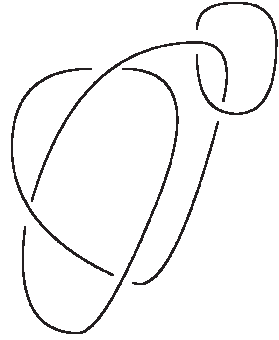
$$}\end{flushleft}
and the manifest formula for the trefoil in the symmetric representations (in the topological framing) \cite[Eqs.(110), (112)]{evo} is
\be
{\cal H}_{[r]}^{3_1} =
1 +  \sum_{s=1}^r \frac{[r]!}{[s]![r-s]!}F_s^{3_1}(A,q)
\prod_{i=1}^s \{Aq^{r+i-1}\}\{Aq^{i-2}\}
\nn
\\
F_s^{3_1}\!(A^2) = q^{s(s-1)/2}A^s \sum_{j=0}^s
(-)^j\frac{[s]!}{[j]![s-j]!}\frac{ \{Aq^{2j-1}\}\cdot(Aq^{j-1})^{2j}}{\prod_{i=j-1}^{s+j-1}\{Aq^i\}}
\nn
\ee

\subsection{L7a3}

\begin{flushleft}
\parbox{10cm}{
This is a 2-component link, made from the trefoil and unknot,
but less trivial than a necklace.
The first representation $r_1$ is for the trefoil and $r_2$, for the unknot. The differences as in (\ref{DElink}) are

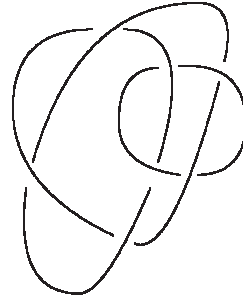
\begin{picture}(300,20)(-320,0)
\qbezier(0,20)(0,50)(30,50)
\qbezier(0,20)(0,-10)(38,-28)
\qbezier(42,50)(70,50)(55,0)
\qbezier(46,-31)(60,-37)(78,30)
\qbezier(5,-10)(0,-50)(25,-50)
\qbezier(7.5,0)(25,60)(70,60)
\qbezier(25,-50)(35,-50)(52,-10)
\qbezier(70,60)(85,60)(80,38)
\qbezier(40,15)(40,35)(55,35.5)
\qbezier(40,15)(40,-2)(63,-5)
\qbezier(63,36)(90,38)(88,10)
\qbezier(70,-5)(85,-5)(88,10)
\end{picture}
\be
\begin{array}{l}
{\cal H}^{L7a3}_{1,1} \ - \ {\cal H}^{3_1}_{1} = \cfrac{A^3 \{q\}^2 [4]\{Aq\}\{A/q\}}{[2]\{A\}}
\\
{\cal H}^{L7a3}_{1,2} \ - \ {\cal H}^{3_1}_{1} = \cfrac{A^3[4]\{q\}^2\{A/q\}\{Aq^2\}}{\{A\}}
\\
{\cal H}^{L7a3}_{1,3} \ - \ {\cal H}^{3_1}_{1} = -\cfrac{A^3[3][4]\{q\}^2\{A/q\}\{Aq^3\}}{[2]\{A\}}\\
{\cal H}^{L7a3}_{2,1} \ - \ {\cal H}^{3_1}_{2} = -\cfrac{A^6[2]\{q\}^2\{A/q\}\{Aq^2\}\Big[A(q^8+q^2)-A^{-1}(q^8+q^4+q^2+q^{-4})\Big]}{\{A\}}\\
{\cal H}^{L7a3}_{2,2} \ - \ {\cal H}^{3_1}_{2} = -\cfrac{A^6[2]^2\{q\}^2 \{Aq^2\}\{A/q\}}{\{A\}\{Aq\}}\sum_{i=-1}^1 \xi_{22}^{(i)}A^{2i}\\
{\cal H}^{L7a3}_{3,1} \ - \ {\cal H}^{3_1}_{3} = -\cfrac{A^9[3]\{q\}^2 \{Aq^3\}\{A/q\}}{\{A\}}\sum_{i=-1}^1 (-1)^{i+1}\xi_{31}^{(i)}A^{2i}\\
{\cal H}^{L7a3}_{2,3} \ - \ {\cal H}^{3_1}_{2} = \cfrac{A^6[2]^2\{q\}^2 \{Aq^3\}\{A/q\}}{\{A\}\{Aq\}}\sum_{i=-1}^1 \xi_{23}^{(i)}A^{2i}\\
{\cal H}^{L7a3}_{3,2} \ - \ {\cal H}^{3_1}_{3} = -\cfrac{A^9[2][3]\{q\}^2 \{Aq^3\}\{A/q\}}{\{A\}\{Aq\}}\sum_{i=-2}^1 \xi_{32}^{(i)}A^{2i+1}\\
{\cal H}^{L7a3}_{3,3} \ - \ {\cal H}^{3_1}_{3} = \cfrac{A^9[3]^2\{q\}^2 \{Aq^3\}\{A/q\}}{\{A\}\{Aq\}\{Aq^2\}}\sum_{i=-2}^2 \xi_{33}^{(i)}A^{2i}\\
\ldots\\
{\cal H}^{L7a3}_{r_1,r_2} \ - \ {\cal H}^{3_1}_{r_1} =\cfrac{A^{3r_1}[r_1][r_2]\{q\}^2 \{Aq^{\max(r_1,r_2)}\}\{A/q\}}{\prod_{i=1}^{\min(r_1,r_2)}\{Aq^{i-1}\}}\Big( \xi_{r_1r_2}^{(-n)}A^{-n}+\xi_{r_1r_2}^{(-n+2)}A^{-n+2}+\ldots+\xi_{r_1r_2}^{(n-2)}A^{n-2}+\xi_{r_1r_2}^{(n)}A^{n}\Big)
\end{array}\nn
\ee}\end{flushleft}
where, in the last line, $n=r_1+r_2-2$ if $r_1\ge r_2$, $n=2(r_1-1)$ if $r_2>r_1$ and
\begin{gather}
\xi_{22}^{(1)}=q^{14}-2q^{12}+3q^{10}-q^{8}-q^{6}+2q^{4}+q^{2}-2+q^{-2},\nn\\
\xi_{22}^{(0)}=-q^{14}+2q^{12}-3q^{10}+q^{6}-4q^{4}+2-3q^{-2}-q^{-4}+2q^{-6}-q^{-8},\ \ \
\xi_{22}^{(-1)}=q^{8}-q^{6}+2q^{4}+q^{-2}+q^{-4}-q^{-6}+q^{-8}\nn\\
\centerline{------------------}\nn\\
\xi_{31}^{(1)}=q^{20}+q^{12},\ \ \
\xi_{31}^{(0)}=q^{20}+q^{18}+q^{14}+2q^{12}+q^{10}+q^4+q^2,\ \ \
\xi_{31}^{(-1)}=q^{18}+q^{14}+q^{12}+q^{10}+q^6+q^4+q^2+q^{-6}\nn\\
\centerline{------------------}\nn\\
\xi_{23}^{(1)}=q^{16}-q^{14}+2q^{10}-2q^8+3q^4-q^2-1+q^{-2},\ \ \ \xi_{23}^{(-1)}=q^8+2q^2-1+2q^{-4}-q^{-8}+q^{-10}\nn\\
\xi_{23}^{(0)}=-q^{16}+q^{14}-2q^{10}+q^8-q^6-3q^4+2-2q^{-2}-2q^{-4}+q^{-6}+q^{-8}-q^{-10}\nn\\
\centerline{------------------}\nn\\
\xi_{32}^{(1)}=q^{28}-q^{26}-q^{24}+3q^{22}-2q^{18}+2q^{14}+q^{12}-q^{10}-q^8+q^6,\ \ \ \xi_{32}^{(-2)}=-q^{18}-2q^{12}-q^6-2q^4-q^{-4}-q^{-6}+q^{-10}-q^{-12}\nn\\
\xi_{32}^{(0)}=-q^{28}+2q^{24}-3q^{22}-3q^{20}+3q^{18}+q^{16}-6q^{14}-3q^{12}+3q^{10}+2q^8-3q^6-3q^4+2-q^{-4}\nn\\
\xi_{32}^{(-1)}=q^{26}-q^{24}+3q^{20}-q^{16}+4q^{14}+4q^{12}-2q^{10}+3q^6+4q^4-2+3q^{-4}+q^{-6}-q^{-8}-q^{-10}+q^{-12}\nn\\
\centerline{------------------}\nn\\
\xi_{33}^{(2)}=-q^{37}+2q^{35}-q^{33}-3q^{31}+5q^{29}-7q^{25}+
2q^{23}+6q^{21}-q^{19}-6q^{17}-q^{15}+3q^{13}+3q^{11}-q^{9}-4q^{7}+q^{5}+2q^{3}-q\nn\\
\xi_{33}^{(1)}=q^{37}-q^{35}-q^{33}+4q^{31}-q^{29}-5q^{27}+7q^{25}+6q^{23}-11q^{21}-3q^{19}+16q^{17}+
5q^{15}-11q^{13}-\nn\\
-6q^{11}+7q^{9}+11q^{7}-q^{5}-9q^{3}-2q+6q^{-1}+3q^{-3}-3q^{-5}-q^{-7}+q^{-9}\nn\\
\xi_{33}^{(0)}=-q^{35}+2q^{33}-q^{31}-4q^{29}+4q^{27}+q^{25}-8q^{23}+2q^{21}+4q^{19}-8q^{17}-9q^{15}+5q^{13}+6q^{11}-8q^{9}-\nn\\
-13q^{7}+q^{5}+10q^{3}+2q-11q^{-1}-6q^{-3}+5q^{-5}+6q^{-7}-2q^{-9}-5q^{-11}+q^{-13}+2q^{-15}-q^{-17}\nn\\
\xi_{33}^{(-1)}=q^{27}-q^{25}+3q^{21}-q^{17}+4q^{15}+4q^{13}-2q^{11}+q^{9}+7q^{7}+3q^{5}-3q^{3}-q+5q^{-1}+5q^{-3}-q^{-5}-4q^{-7}+\nn\\
+5q^{-11}+q^{-13}-3q^{-15}+q^{-19}\nn\\
\xi_{33}^{(-2)}=-q^{17}+q^{15}-q^{13}-2q^{11}+q^{9}-3q^{5}+q-q^{-1}-2q^{-3}+q^{-9}-q^{-11}-2q^{-13}+q^{-15}+q^{-17}-q^{-19}\nn
\end{gather}

\bigskip

This is also in accordance with the prediction of (\ref{Nrefll2}) and (\ref{Nrefl2}):
\be
{\cal H}^{L7a3}_{r_1,r_2} - {\cal H}^{L7a3}_{N-r_1,N-r_2} \ \vdots \
\{Aq^N\}\{A/q\}
\ee
$N=r_1\geq r_2$:
\be
{\cal H}^{L7a3}_{r_1,r_2} - {\cal H}^{L7a3}_{0,r1-r_2}
= {\cal H}^{L7a3}_{r_1,r_2} -1 \ \ \ \vdots \ \ \
\{Aq^{r_1}\}\{A/q\}
\ee
$N=r_2\geq r_1$:
\be
{\cal H}^{L7a3}_{r_1,r_2} - {\cal H}^{L7a3}_{r_2-r_1,0}
=  {\cal H}^{L7a3}_{r_1,r_2} - {\cal H}^{3_1}_{r_2-r_1}  \ \ \  \vdots \ \ \
\{Aq^{r_2}\}\{A/q\}
\ee

\section{Conclusion}

In this letter, we discovered the structure of differential expansion for links.
We did this only for symmetric representations
(and antisymmetric obtained by the substitution $q\to -1/q$),
still this is non-trivial.
The framing providing the differential expansion, i.e. respecting the equivalence
of link polynomials in representations and their conjugates
appeared to be different from the canonical one typically used for links.
Better understanding of this new discovery and its extension to non-symmetric
representations are the next steps to do.

\section*{Acknowledgements}

The work was partly supported by the grant of the Foundation for the Advancement of Theoretical Physics ``BASIS" (A.Mor. and A.S.), by grant MK-8769.2016.1 (A.S.), by RFBR grants 16-01-00291 (A.Mir.), 16-02-01021 (A.Mor.), 17-01-00585 (An.Mor.) and 16-31-60082-mol-a-dk (A.S.), by joint grants 17-51-50051-YaF, 15-51-52031-NSC-a, 16-51-53034-GFEN, 16-51-45029-IND-a (A.M.'s and A.S.) and by grant NSFC (11425104, 11611130015) (C.B., J.J. and J.L.). Chengming Bai, Jianjian Jiang and Jinting Liang thank ITEP for the hospitality and for useful discussions while they visited there in the summer of 2017.

\end{document}